\documentclass[conference]{IEEEtran}
\IEEEoverridecommandlockouts
\usepackage{fancyhdr}

\usepackage{cite}
\usepackage{amsmath,amssymb,amsfonts}
\usepackage{graphicx}
\usepackage{textcomp}
\usepackage{xcolor}
\usepackage{float}

\usepackage{multirow}   %表格
\usepackage{makecell}   %表格
\usepackage{booktabs}   % 表格
\usepackage[noend]{algpseudocode} %伪代码
\usepackage{algorithmicx,algorithm} %算法
\usepackage{wrapfig}

\def\BibTeX{{\rm B\kern-.05em{\sc i\kern-.025em b}\kern-.08em
    T\kern-.1667em\lower.7ex\hbox{E}\kern-.125emX}}

\begin{document}

\title{	PIMCOMP: A Universal Compilation Framework for Crossbar-based PIM DNN Accelerators 
\vspace{-0.4cm}
}

\author{
\IEEEauthorblockN{Xiaotian Sun, Xinyu Wang, Wanqian Li, Lei Wang, Yinhe Han, Xiaoming Chen\IEEEauthorrefmark{1}}
\IEEEauthorblockA{Center for Intelligent Computing Systems, Institute of Computing Technology, Chinese Academy of Sciences}
\IEEEauthorblockA{University of Chinese Academy of Sciences}
\IEEEauthorblockA{\IEEEauthorrefmark{1}Corresponding author: chenxiaoming@ict.ac.cn}
\vspace{-1.3cm}
}

\maketitle
\thispagestyle{fancy}
\chead{This paper is published in Design Automation Conference 2023 (DAC'23)} %指定页眉中间内容
\cfoot{}
% \lfoot{979-8-3503-2348-1/23/\$31.00~\copyright~2023~IEEE}
% \cfoot{}
\renewcommand{\headrulewidth}{0mm}

\begin{abstract}
Crossbar-based PIM DNN accelerators can provide massively parallel in-situ operations. A specifically designed compiler is important to achieve high performance for a wide variety of DNN  workloads. However, some key compilation issues such as parallelism considerations, weight replication  selection, and array mapping methods have not been solved. In this work, we propose PIMCOMP - a universal compilation framework for NVM crossbar-based PIM DNN accelerators. PIMCOMP is built on an abstract PIM accelerator architecture, which is compatible with the widely used Crossbar/IMA/Tile/Chip hierarchy. On this basis, we propose four general compilation stages for crossbar-based PIM accelerators: node partitioning, weight replicating, core mapping, and dataflow scheduling.  We design two compilation modes with different inter-layer pipeline granularities to support high-throughput and low-latency application scenarios, respectively. Our experimental results show that PIMCMOP yields improvements of 1.6$\times$ and 2.4$\times$ in throughput and latency, respectively, relative to PUMA.
\end{abstract}

\begin{IEEEkeywords}
NVM, PIM accelerator, deep neural network, compilation framework
\end{IEEEkeywords}

\setlength\intextsep{-0.2pt} %浮动体底部与后面正文的距离

\vspace{-0.2cm}
\section{Introduction}
\vspace{-0.15cm}
In recent years, deep neural networks (DNNs) have made breakthroughs in a variety of tasks. Due to the expansion of the parameter scale of DNN models, the industry expects considerable performance improvements in hardware to efficiently run DNN algorithms. Therefore various DNN accelerators (e.g., \cite{Diannao, Eyeriss}) have been proposed. However, these devices based on CMOS and following the von Neumann architecture are now hitting the memory wall challenge \cite{MemoryWall}, and encounter bottlenecks in the improvements of storage, bandwidth, and energy consumption.

% Process-in-memory (PIM) is regarded as an important technology to avoid the memory wall, and it has become a popular research direction in the field of DNN accelerators. PIM has a variety of implementations, among which emerging Non-Volatile Memory (NVM) devices such as RRAM \cite{RRAMWong}, PCM \cite{PCM1}, MRAM \cite{MRAM2} have great potential to challenge the dominance of CMOS. Integrating these devices into a 2-D cross-point array results in a crossbar array. The 2-D crossbar structure formed by NVM devices has attracted growing interest due to its high memory density and parallel in-situ computing properties \cite{ReRAMApp}.

Process-in-memory (PIM) is regarded as a promising technology to avoid the memory wall, and has become a popular research direction in the field of DNN accelerators. PIM has a variety of implementations, among which emerging Non-Volatile Memory (NVM) devices such as RRAM, PCM and MRAM have great potential to challenge the dominance of CMOS. Integrating these devices into a 2D cross-point array results in a crossbar array. The 2D crossbar structure formed by NVM devices has attracted growing interest due to its high memory density and parallel in-situ computing properties \cite{ReRAMApp}.

There are previous works (e.g., \cite{ISAAC, AtomLayer, Brahms, MRAM, PCM2}) proposing NVM crossbar-based PIM DNN accelerators. However, they mainly focus on the design of specific architectures and lack consideration of the execution details of DNNs. First, existing works rely on manually mapping the weight data to the crossbars, which ignores the impact of weight mapping on the parallelism of the crossbars and has poor scalability. Second, existing designs often determine the replication of the weights intuitively, such as replicating weight data in early layers to keep the execution pipeline balanced. However, this method does not make effective use of resources. It needs a toolchain to elaborate the task mapping, resource allocation, data distribution, dataflow scheduling, etc., to fully exploit the performance of PIM DNN accelerators when running DNN models. Although PUMA \cite{PUMA} has proposed a compiler for memristor crossbar-based accelerators, it has not effectively solved the problems mentioned above.

To address these limitations, we propose a universal compilation framework, PIMCOMP, for NVM crossbar-based PIM accelerators. The compilation process is divided into 4 phases: node partitioning, weight replicating, core mapping, and dataflow scheduling. Based on these steps, we design a DNN inference compiler.  We make the following contributions.
\begin{itemize}
    \item We propose a general and representative NVM crossbar based accelerator architecture as a hardware abstraction for studying the actual execution of DNNs;

    \item We provide two compilation modes: ``low latency'' and ``high throughput'', according to the user's application scenario;

    \item We propose a genetic algorithm to optimize weight replication and core mapping, with novel fitness function for both compilation modes;

    \item We design scheduling algorithms to support DNN networks with complex topologies and multiple operators and optimize on-chip memory usage.
\end{itemize}

\section{Background and Motivation}

\subsection{NVM Crossbar based DNN Accelerators}

% \begin{figure}[htbp]
%   \centering
% %   \vspace{-0.5cm} % 调整图片与上文的垂直距离
%   \setlength{\abovecaptionskip}{-0.1cm} %调整标题上方的距离   
%   \setlength{\belowcaptionskip}{-0.05cm} %调整标题下方的距离 	   
%   \setlength{\abovedisplayskip}{0pt} %与上方展示内容的距离	
%   \setlength{\belowdisplayskip}{0pt} %与下方展示内容的距离
%   \includegraphics[width=0.4\linewidth]{fig/Xbar2.pdf}
%   \caption{A NVM crossbar with peripheral devices}
%   \label{fig::xbar}
% % \vspace{-0.5cm}
% \end{figure}

\begin{wrapfigure}{r}{0.18\textwidth}
% \hspace{2.5cm}
%   \centering
%   \vspace{-0.5cm} % 调整图片与上文的垂直距离
  \setlength{\abovecaptionskip}{-0.2cm} %调整标题上方的距离   
  \setlength{\belowcaptionskip}{-0.2cm} %调整标题下方的距离 	   
  \includegraphics[width=\linewidth]{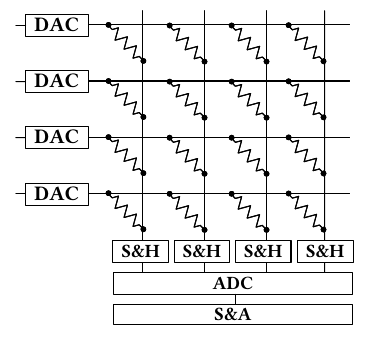}\\
  \caption{A NVM crossbar with peripheral devices}
  \label{fig::xbar}
\end{wrapfigure}

\vspace{-0.1cm}
Fig. \ref{fig::xbar} shows the structure of the NVM crossbar which is the basic operation unit of PIM DNN accelerators. Due to hardware characteristics, NVM crossbars can perform matrix-vector multiplication (MVM) operations efficiently in an analog manner. The conductance of a crossbar cell at each cross-point can be programmed to store an element of matrix $\textbf{G}$ and a voltage vector representing the vector $\textbf{V}$ is applied to the rows. According to Ohm's law, the current value at each cross-point is $I_{ij} = G_{ij}V_j$, and according to Kirchhoff's law, the accumulated current value that can be read in each column is $I_i=\sum_j G_{ij }V_j$. In this way, the crossbar can calculate the dot product of  matrix $\textbf{G}$ and vector $\textbf{V}$ in parallel. Since the NVM crossbar operates in the analog domain and other circuits of the accelerator work in the digital domain, peripheral circuits such as ADC/DAC, Sample and Hold (S\&H), Shift And Add (S\&A) are required in the operation unit.

To speed up  DNN inference, the weight data is mapped into the crossbar, and the input is fed as the voltage input. So the crossbar can complete the MVM operations in DNNs in parallel and previous works (e.g., \cite{ISAAC, MRAM, PCM2}) have used NVM crossbars to accelerate DNN inference.

% ISAAC \cite{ISAAC} is a full-ﬂedged CNN accelerator based on RRAM crossbar arrays. In ISAAC, the RRAM crossbars are organized in a "Chip-Tile-IMA" hierarchy. Inter-layer pipelines are implemented to reduce on-chip cache requirements. The balance of the pipeline is achieved by duplicating the weights of convolutional layer. A coding method is proposed to solve the problem of positive and negative weights and reduce ADC/DAC overhead. In addition, there are many other works based on RRAM device \cite{PRIME, AtomLayer}.

% Due to the small resistance value of MRAM, the current accumulation value in the MRAM crossbar is too large, resulting in higher power consumption. To solve this problem, Jung et al. \cite{MRAM} proposed a resistor accumulation method to reduce the power consumption of analog multiply–accumulate operations. Chen et al.\cite{PCM2} solved the inherent non-ideal characteristics of PCM such as write noise, variability, and conductance drift to a certain extent by improving programming accuracy and introducing slope correction technology, and used PCM crossbar array for high-performance inference accelerators. 

% Although previous works choose different NVM materials, they share a common goal to accelerate MVM operations. However, the previous works focus on the design of the architecture and the details of the DNN network operation process have not been paid attention to, such as the parallelism between NVM crossbars and the scheduling method of complex networks.

\vspace{-0.1cm}
\subsection{DNN Compilation Framework}
\vspace{-0.05cm}

To bridge the gap between various Deep Learning (DL) frameworks and different DL hardwares, several DNN compilation frameworks have been proposed by industry and academia (e.g., \cite{TVM, TC}). 

%The DNN compilation framework takes the DNN model and parameters as input, performs multi-stage optimization, and finally obtains code suitable for running on different hardware.

TVM \cite{TVM} is a popular end-to-end DL optimization stack that provides multi-stage back-end optimization
for different hardwares. However, TVM is not suitable for PIM architectures. First,  storage units in PIM are also computation units, which makes the storage and computing characteristics of PIM different from traditional hardware. Second, a significant stage in TVM's optimization of computing is the scheduling of multi-layer loops, including unrolling, vectorization, parallelization, and tiling. However, the parallel MVM operation is naturally supported by PIM crossbars, so TVM optimization will have limited effects on NVM crossbar-based accelerators.

In the field of PIM, PUMA \cite{PUMA} is the first memristor-based ML inference accelerator that supports ISA with a compiler that can convert high-level languages into ISA code. Nonetheless, heuristic weight replicating and core mapping methods adopted by its compiler are difficult to guarantee high performance. In addition, the granularity of PUMA's inter-layer pipeline is inference, that is, different layers process data of different inferences, and this processing manner is unacceptable in low-latency scenarios.

% \vspace{-0.15cm}
\section{Abstract Accelerator Architecture}

% \vspace{-0.1cm}
We propose an abstract PIM accelerator architecture. On this basis, we propose an execution model to capture the execution characteristics and parallelism of PIM accelerators.

\vspace{-0.1cm}
\subsection{Hardware Abstraction}

\begin{figure}[b]
  \centering
  \vspace{-0.3cm} % 调整图片与上文的垂直距离
  \setlength{\abovecaptionskip}{-0.1cm} %调整标题上方的距离   
  \setlength{\belowcaptionskip}{-0.05cm} %调整标题下方的距离 	   
  \setlength{\abovedisplayskip}{-5pt} %与上方展示内容的距离	
  \setlength{\belowdisplayskip}{0pt} %与下方展示内容的距离
  \includegraphics[width=0.85\linewidth]{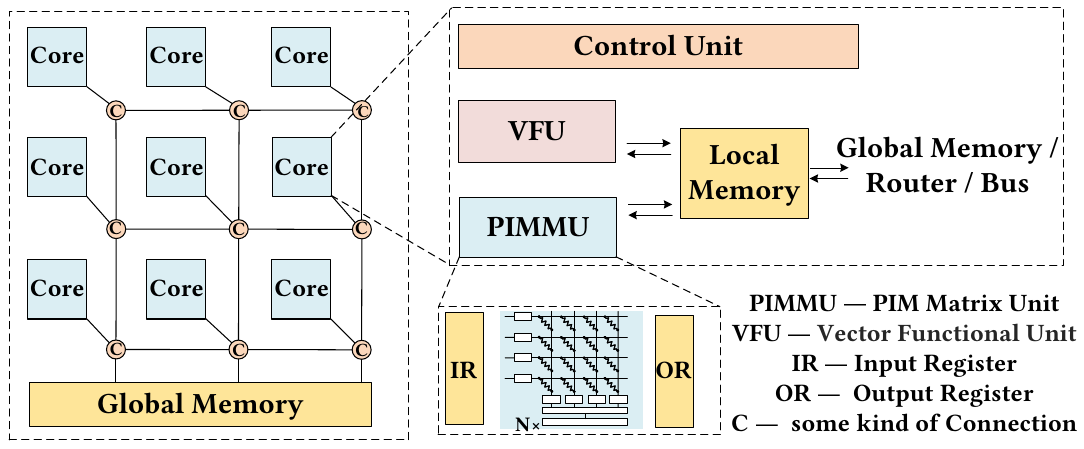}\\
  \caption{Abstract PIM accelerator architecture}
  \label{fig::arch}
% \vspace{-0.05cm}
\end{figure}

Fig. \ref{fig::arch} illustrates the proposed abstract DNN accelerator architecture. At a high level, the accelerator consists of a series of cores connected to a global memory. Each core can be controlled by instructions or state machines. The cores can be interconnected through NoC or busses. The weights of the neural network are stored in the cores, while the inputs, outputs and intermediate results are stored in the global memory. Different cores perform operations asynchronously. Inter-core synchronization occurs when there is an inter-core data transfer.

% \begin{figure*}
% \begin{minipage}[t]{0.5\textwidth}%并排放两张图片，每张占页面的0.5，下同。
% \centering
% \includegraphics[width=\textwidth]{fig/Overview.pdf}
% \caption{The Overview of PIMCOMP}
% \label{fig::overview}
% \end{minipage}
% \begin{minipage}[t]{0.45\textwidth}
% \centering
% \includegraphics[width=\textwidth]{fig/Weight2.pdf}
% \caption{Node Partitioning Strategy}
% \label{fig::weight}
% \end{minipage}
% \vspace{-0.3cm}
% \end{figure*}

The abstract core in Fig. \ref{fig::arch} consists of a control unit, a local memory and two types of operation units: PIM matrix unit (PIMMU) and vector functional unit (VFU). The PIM matrix unit is composed of several PIM crossbars, which are used to perform MVM operations in the DNN, while the VFU performs other operations, including activation function, pooling, and element-wise operations. The local memory works as a scratchpad that stores inputs and outputs of DNN nodes. The PIMMU and VFU can only access data in the local memory. In addition, operations such as padding, concatenation, and split can also be handled using the local memory. 

Our proposed abstract architecture is compatible with the Crossbar/IMA/Tile/Chip structure widely adopted in previous work \cite{ISAAC}. This work focuses on the general optimization idea of the compilation process of crossbar-based PIM DNN accelerator, and we do not consider the detailed optimization inside the PIM matrix unit. Nonetheless, related optimizations such as mixed size crossbars \cite{MIXED} and low-bit ADCs \cite{InfoX} are compatible with this abstract architecture.

%Here, the core and tile are in the consistent level. The PIM matrix unit in core can be considered to be composed of a series of IMAs \cite{ISAAC}, while other functional components and eDRAM Buffer in the tile \cite{ISAAC} can be represented by vector functional unit and local memory, respectively.

% \begin{figure*}[h!]
%   \centering
% %   \vspace{-0.5cm} % 调整图片与上文的垂直距离
%   \setlength{\abovecaptionskip}{-0.1cm} %调整标题上方的距离   
%   \setlength{\belowcaptionskip}{-0.05cm} %调整标题下方的距离 	   
%   \setlength{\abovedisplayskip}{0pt} %与上方展示内容的距离	
%   \setlength{\belowdisplayskip}{0pt} %与下方展示内容的距离
%   \includegraphics[width=0.8\linewidth]{fig/Overview2.pdf}
%   \caption{The Overview of PIMCOMP}
%   \label{fig::overview}
% \end{figure*}

\vspace{-0.06cm}
\subsection{Execution Model}
\vspace{-0.08cm}

According to our abstract architecture, after mapping the DNN to the cores, each core can get a static operation sequence, which is composed of the basic operations such as MVM, VEC, COMM, and MEM, representing MVM operations by PIMMU, vector operations by VFU, communication between cores, and access to global memory, respectively. %The communication between cores is performed in a blocking manner. Only after the send and receive operations match, the two cores can continue to run the following operations.
We do not restrict the format of the operation sequence. It can be a series of instructions, or a schedule of basic operators, etc.

For any two MVMs in the operation sequence of the same core, the parallel relationship between them is analyzed as follows.
\begin{itemize}
    \item If they are for the same crossbar, that is, there is a structural conflict between them, the latter one must wait for the previous one to complete before it can start.
    
    \item If  the input data of the latter MVM is the output of the former one, that is, there is a data dependency between them, the latter one must also wait for the former MVM to finish before it can start.
\end{itemize}

% If there is a COMM operation between these two MVM operations, the start time of the latter MVM must be after the COMM operation is completed;

For several MVM operations that do not have structural conflicts, data dependencies, and synchronization blocking at a certain moment, the start time interval between two adjacent MVM operations is determined by the on-chip bandwidth of each core.

% For several MVM operations that do not have structural conflicts, data dependencies, and synchronization blocking at a certain moment, let the start time interval of two adjacent MVM operations be $t_{interval}$. $t_{interval}$ is determined by the on-chip bandwidth of each core.

% In order to support the PIM Matrix Unit, the minimum on-chip bandwidth requirement is $BW_{min}=\frac{V_{read}}{T_{read}}$, where $V_{read}$ is the amount of data required by one crossbar array to perform a complete operation and $T_{read}$ is the time;

% If the PIM Matrix Unit does not support the parallel operation of the crossbar, then $t_{int} = t_{MVM}$;

% If the PIM Matrix Unit supports the parallel execution, assuming that the on-chip bandwidth is $BW$, then $t_{int} = \frac{T_{read}}{ \lfloor \frac{BW}{BW_{min} } \rfloor}$

% From this, the execution data flow of one core can be determined.

% \vspace{-0.1cm}
\section{Compilation Framework}

\subsection{The Overview of PIMCOMP}

\begin{figure}[htbp]
  \centering
%   \vspace{-0.5cm} % 调整图片与上文的垂直距离
  \setlength{\abovecaptionskip}{-0.1cm} %调整标题上方的距离   
  \setlength{\belowcaptionskip}{-0.05cm} %调整标题下方的距离 	   
  \setlength{\abovedisplayskip}{0pt} %与上方展示内容的距离	
  \setlength{\belowdisplayskip}{0pt} %与下方展示内容的距离
  \includegraphics[width=\linewidth]{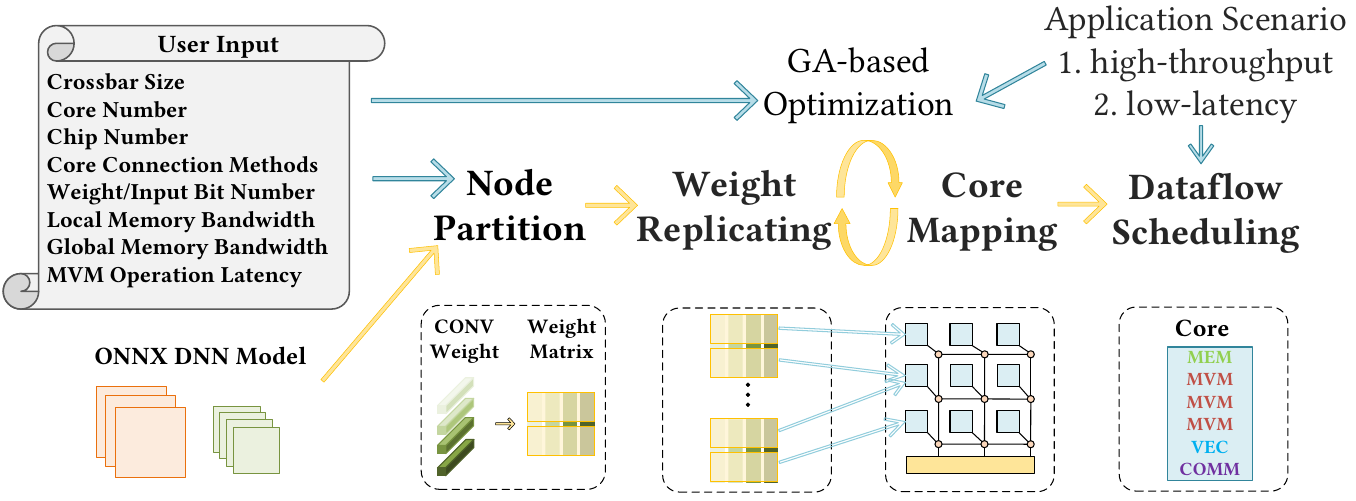}\\
  \caption{The overview of PIMCOMP}
  \label{fig::overview}
\vspace{-0.7cm}
\end{figure}

Fig. \ref{fig::overview} illustrates the high-level overview of PIMCOMP. First, PIMCOMP reads the user input and loads DNN model in ONNX format which facilitates conversion between different DL frameworks, and obtains the model description including node (in this work, node and layer share the same meaning) information and topological relationship after parsing the model. This information will be fed to the backend. The entire backend includes 4 stages. Node Partitioning (Section \ref{sec::node_pratitioning}) describes the rules for dividing the convolutional and fully connected layers according to the crossbar size. Weight Replicating (Section \ref{sec::replicating_and_mapping}) determines the replication numbers of different nodes, Core Mapping (Section \ref{sec::replicating_and_mapping}) decides the mapping relationship between crossbars and cores, and Dataflow Scheduling (Section \ref{sec::dataflow_scheduling}) performs scheduling and optimization according to user's requirement to generate control flow or instruction flow.

We provide two compilation modes for users to choose from: High Throughput (HT) and Low Latency (LL), which are suitable for scenarios with continuous input data of large batches and intermittent input of a small amount of data, respectively. Their main design difference is the granularity of the inter-layer pipeline. In HT mode, DNN processes in a layer-by-layer manner. When the pipeline is filled, different layers process data from different inferences. There is no inter-layer data communication so parallelism between layers is high. In LL mode, as long as one layer produces an output, it passes the output to the position required by subsequent layers. When a layer receives enough data, it can start its operation, so the overall latency is low. 
% Inter-layer information transfer occurs whenever an output is generated. The data dependency between layers is strong, and it is difficult to take advantage of the parallelism within the core, so the throughput is low.

\vspace{-0.1cm}
\subsection{Node Partitioning}
\label{sec::node_pratitioning}

\begin{figure}[htbp]
  \centering
  \vspace{-0.4cm} % 调整图片与上文的垂直距离
  \setlength{\abovecaptionskip}{-0.1cm} %调整标题上方的距离   
  \setlength{\belowcaptionskip}{-0.05cm} %调整标题下方的距离 	   
  \setlength{\abovedisplayskip}{0pt} %与上方展示内容的距离	
  \setlength{\belowdisplayskip}{0pt} %与下方展示内容的距离
  \includegraphics[width=0.9\linewidth]{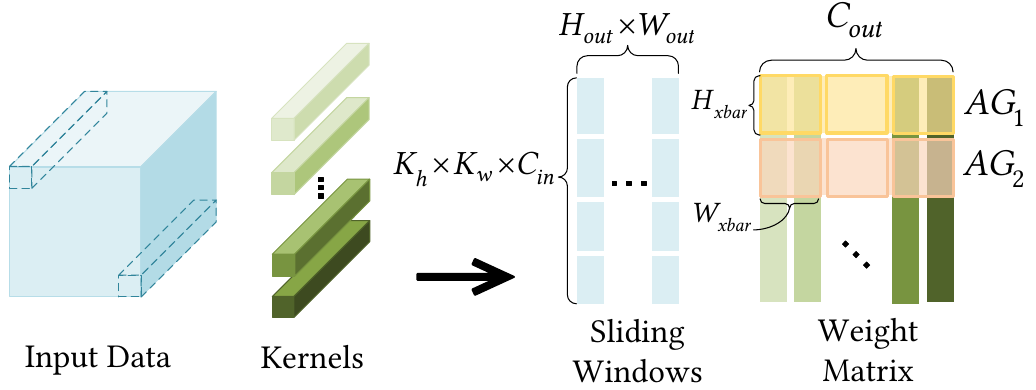}\\
  \caption{Node Partitioning strategy}
  \label{fig::weight}
\vspace{-0.0cm}
\end{figure}

\textcolor{black}{Due to the limited size of a crossbar array, weight data of convolutional layers and fully connected layers cannot be completely mapped to the same crossbar in general, so we need to partition these nodes to fit in the crossbars.} Node partitioning strategy is shown in Fig. \ref{fig::weight}. First, the convolutional layers and fully connected layers (which can be regarded as special convolutional layers) in DNN  are converted into MVM operations. Specifically, the weights of each convolution kernel are flatten into a column, thus obtaining a weight matrix with height $k_w \times k_h \times C_{in}$ and width $C_{out}$, where $k_w$ , $k_h$ represent the width and height of the convolution kernel and $C_{in}$, $C_{out}$ represent the number of input channels and output channels, respectively. After that, the entire weight matrix is divided into several Array Groups (AG) according to the size of the crossbar array. The height of each AG is equal to the height of the crossbar array $H_{xbar}$, and the width is equal to $C_{out}$. Therefore each AG contains $\lceil C_{out}/W_{xbar} \rceil$ crossbar arrays, where $W_{xbar}$ is the width of crossbar array. Each AG needs to run $H_{out} \times W_{out}$ input cycles (input sliding windows), where $H_{out}$ and $W_{out}$ are the height and width of the output feature.

It is preferred to map all crossbars of one AG to the same core. Crossbar arrays belonging to the same AG can be driven by the same instruction or control signal, therefore gathering them in one core can reduce control complexity. More importantly, these crossbar arrays have the exact same input. If they are in the same core, input data can be broadcast to these arrays, which avoids repeatedly accessing Input Register and alleviates on-chip bandwidth and buffer pressure.

After partitioning the weight, the AGs need to be allocated to the cores. One core may contain AGs of multiple nodes, and AGs of one node may be mapped to different cores. The MVM results obtained by AGs of the same node need to be accumulated to get a complete computation result. If AGs of the same node are mapped to different cores, data accumulation across cores is required.

% \vspace{-0.2cm}
\subsection{Weight Replicating and Core Mapping}
\label{sec::replicating_and_mapping}

% The number of sliding windows of each replicated weight block will be correspondingly reduced.
\vspace{-0.1cm}
\textcolor{black}{The storage units in PIM are also computation units, so an important way to improve computation parallelism is to replicate the weight data multiple times. Besides, allocating computing tasks among cores also greatly influences performance of PIM accelerators, taking into account structural conflict and data dependency. The above two steps are intertwined so we employ a modified genetic algorithm to optimize them simultaneously.}

\subsubsection{Algorithm Design}

In the genetic algorithm, each gene represents several AGs of a node, which is encoded as an integer, expressed as $node\_index \times 10000 + AG\_num$. For example, $1030025$ represents $25$ AGs of the $103rd$ node. To ensure that the mapping result is not so scattered that on-chip communication becomes a bottleneck,  we set a limit on the number of nodes that each core can hold, $max\_node\_num\_in\_core$. Therefore, the chromosome length of each individual is $core\_num \times max\_node\_num\_in\_core$, where $core\_num$ is user specified. The location of each gene in the chromosome determines the index of core these AGs are mapped to. This encoding takes into account flexibility and operational efficiency. If each AG is coded, the chromosome will be too long to process efficiently.

In the initialization phase, we randomly select the replication number for each node, and randomly map the AGs to the cores. The design of the fitness function is directly related to the optimization effect, which will be described in the following subsection. Crossover phase in this issue lacks practical significance, so we skip it. Mutation is an important phase to improve resource utilization and inference performance. In this stage, the algorithm will randomly select individuals to perform one of the following four mutation operations: \textbf{\MakeUppercase{\romannumeral 1}}. Randomly select a node, increase its replication number, and randomly map the expanded AG to cores. \textbf{\MakeUppercase{\romannumeral 2}}. Randomly select a node, reduce its replication number, and recover the crossbar arrays occupied by that replicated block. \textbf{\MakeUppercase{\romannumeral 3}}. Randomly select a gene and spread its AG to other cores. \textbf{\MakeUppercase{\romannumeral 4}}. Randomly select a gene and merge its AG into the same node of other cores.

% \begin{itemize}
%     \item Randomly select a node, increase its replication number, and randomly map the expanded AG to cores.
%     \item Randomly select a node, reduce its replication number, and recover the crossbar arrays occupied by that replicated block.
%     \item Randomly select a gene and spread its AG to other cores. 
%     \item Randomly select a gene and merge its AG into the same node of other cores.
% \end{itemize}

\subsubsection{Fitness Function}

\begin{figure}[htbp]
\vspace{-0.2cm}
  \centering
%   \vspace{-0.5cm} % 调整图片与上文的垂直距离
  \setlength{\abovecaptionskip}{-0.15cm} %调整标题上方的距离   
  \setlength{\belowcaptionskip}{-0.15cm} %调整标题下方的距离 	   
  \setlength{\abovedisplayskip}{0cm} %与上方展示内容的距离	
  \setlength{\belowdisplayskip}{0pt} %与下方展示内容的距离
  \includegraphics[width=\linewidth]{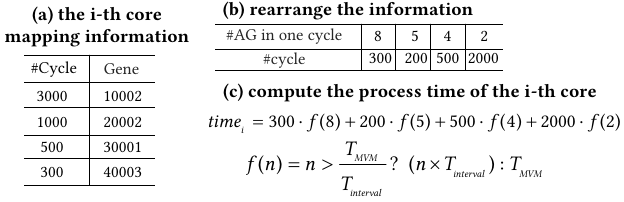}\\
  \caption{An example to estimate execution time of the i-th core}
  \label{fig::fitHT}
% \vspace{-0.5cm}
\end{figure}

\textbf{For HT mode}, the overall inference time can directly reflect the performance. Fig. \ref{fig::fitHT} is an example of how to calculate the estimated inference time of the i-th core in HT mode. A total of 4 nodes are mapped to the i-th core, containing $2$, $2$, $1$, and $3$ AGs respectively, and the AGs of each node have $3000$, $1000$, $500$, and $300$ input sliding windows respectively. Because there is no data dependency between AGs, each AG starts to execute in turn at interval $T_{interval}$ under the condition that no structural conflict occurs. Rearrange the information to get the table in Fig. \ref{fig::fitHT}(b). The $(8, 300)$ in the first column means that the number of AGs in this core is $8$ for the first $300$ operation cycles. After $300$ cycles, the 4-th node is completed, so the number of AGs for the next $200$  cycles is $5$. Based on this information, the estimated time of the i-th core can be calculated according to the formula in Fig. \ref{fig::fitHT}(c). $f(n)$ calculates the operation time of one operation cycle when there are $n$ AGs in the core. If $n>\frac{T_{MVM}}{T_{interval}}$, where $T_{MVM}$ is the time to complete a single MVM operation, then each operation cycle takes $f(n)=n \times T_{interval}$; otherwise $f(n)=T_{MVM }$. In this way, an ideal inference time $time_i$ is calculated for each core, so the fitness function for HT mode is $F_{HT} = \max_i{time_i}$.

\begin{figure}[htbp]
  \centering
%   \vspace{-0.5cm} % 调整图片与上文的垂直距离
  \setlength{\abovecaptionskip}{-0.1cm} %调整标题上方的距离   
  \setlength{\belowcaptionskip}{-0.05cm} %调整标题下方的距离 	   
  \setlength{\abovedisplayskip}{0pt} %与上方展示内容的距离	
  \setlength{\belowdisplayskip}{0pt} %与下方展示内容的距离
  \includegraphics[width=\linewidth]{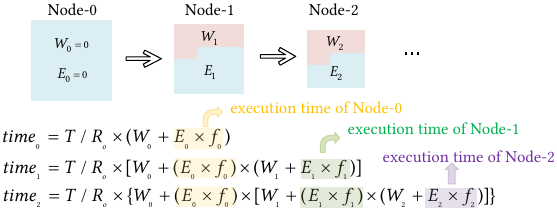}\\
  \caption{An illustration to estimate runtime of LL mode}
  \label{fig::fitLL}
% \vspace{-0.6cm}
\end{figure}

% \textbf{For LL mode}, the inter-layer execution sequence in LL mode can be equivalent to ``after waiting for the provider node (layer) to generate enough output, the consumer node (layer) starts to execute and generates all outputs without pause''. This waiting percentage $W$ can be calculated for each node. Further, if the uninterrupted execution time of the provider node is $T_p$ and if the replication number of the provider node is $r$ times that of the consumer, it takes $T_p \times r \times (1-W)$ for the consumer to complete the calculation without pausing after waiting $T_p \times W$. 
% Therefore the overall time to compute these two nodes is $T_p \times ( W +  r \times (1-W))$.

\textbf{For LL mode}, the inter-layer execution sequence in LL mode can be equivalent to ``after waiting for the provider node (layer) to generate enough output, the consumer node (layer) starts to execute and generates all outputs without pause''. This waiting percentage $W$ can be calculated for each node. Further, if the uninterrupted execution time of node $m$ is $T_m$ and if the replication number of node $m$ is $r$ times that of its consumer node $n$, it takes $T_m \times r \times (1-W_n)$ for the consumer node $n$ to complete the calculation without pausing after waiting $T_m \times W_n$. 
Therefore the overall time to complete computing these two nodes is $T_m \times ( W_n +  r \times (1-W_n))$.

The estimated runtime of LL mode can be derived as shown in Fig. \ref{fig::fitLL}. For convenience, we set $f_x = \min (\frac{R_{p(x)}}{R_{x}},1)$, where $R_x$ is the replication number of the node $x$ and $p(x)$ gets the provider node index of node $x$. We let $E_x=1-W_x$, which represents the percentage of execution of node $x$. We assume the execution time of the first node without extra replication is $T$. Iterate based on topology and the final estimated time is used as the fitness function $F_{LL}$.

\vspace{-0.3cm}
\subsection{Dataflow Scheduling}
\vspace{-0.2cm}

\label{sec::dataflow_scheduling}

The dataflow scheduling stage will generate a sequence of instructions or control flow according to the pipeline mode selected by the user. We first introduce the dataflow scheduling algorithms in HT mode and LL mode, and then describe the on-chip memory optimization technique.

% The convolutional layer and fully connected layer rely on the PIM Matrix Unit to perform, and other operations are called post operations, which will be processed by VFU or local memory.

\vspace{0.2cm}
\subsubsection{HT Dataflow}  
\begin{algorithm}[htbp]
\caption{HT Dataflow Scheduling Algorithm} %算法的名字
\label{alg::HT}
% \hspace*{0.02in} {\bf Input:} %算法的输入
% input parameters A, B, C \\
% \hspace*{0.02in} {\bf Output:} %算法的结果输出
% output result
\begin{algorithmic}[1]
\For{each core}
    \While{have unfinished AG}
        \State load data from global memory
        \For{each unfinished AG}
            \State perform one MVM operation
        \EndFor    
        \State accumulate results across AGs within core
        \State accumulate results across AGs between cores
        \State apply activation function to the results
        \State store data to global memory
    \EndWhile
\EndFor
% \State \Return result
\State allocate other operations to cores
\end{algorithmic}
\end{algorithm}
\vspace{0.1cm}
%%%%%%%%%%%%%%%%%%%%%%%%%%%%%%%%%%%%%%%%%%%%%%%
% 这里要对HT dataflow的生成有更细致的描述。比方说我们选定哪些核进行数据的数据。核内和跨核数据Acc分别怎么做。在哪些核上进行Act。
%%%%%%%%%%%%%%%%%%%%%%%%%%%%%%%%%%%%%%%%%%%%%%%

Algorithm \ref{alg::HT} shows dataflow scheduling for HT mode. It is unrealistic to store all data in a inference of each node in on-chip local memory so it is necessary to periodically transfer some input (output) data from (to) the global memory to (from) the local memory (Lines 3 and 9). In the inter-core data accumulation process (Line 7), AGs are required to transfer the data to be accumulated to the core where the first AG of this replicated weight block is located. To improve parallelism, other operations such as POOL, CONCAT, ELTWISE are distributed among several cores (Line 10). DNNs with complex topology are easy to implement in HT mode with  each node reading  the corresponding data from global memory.

\subsubsection{LL Dataflow}  

In LL mode, each node computes an output and then immediately passes it to its consumer nodes. When one node receives enough input, it can start executing. The condition for the output $(r,c)_i$  of node $i$ to start computing is that the node has received the last input $(r_d,c_d)_i$ \cite{MNSIM2} that it requires, and  $(r_d,c_d)_i$ can be formulated as:

\begin{equation}  
\small
\setlength{\abovedisplayskip}{-5pt}
\setlength{\belowdisplayskip}{-0pt}
(r_{d})_i = 
\left\{  
             \begin{array}{lr}  
             \min \big( H_i, \left( K_i + s_i \times (r-1) - p_i \right) \big) \ \ \   CONV, POOL  \\
             
             H_i \ \ \ \ \ \  \ \ \ \ \ \ \ \ \ \ \  \ \ \ \ \ \ \ \ \ \ \ \ \ \ \ \ \ \ \ \ \ \ \ \ \ \ \ \ \ \ \ \ \ \ \ \ \ \ \ \ \  FC \\
             
             (r_{d})_i \ \ \ \ \ \ \ \ \ \ \ \ \ \ \ \ \ \ \ \ \ \ \ \ \ \ \ \ \  CONCAT, ELTWISE \\
             \end{array}   
\right.  
\nonumber 
\end{equation}  

\begin{equation}  
\small
\setlength{\abovedisplayskip}{-0pt}
\setlength{\belowdisplayskip}{3pt}
(c_{d})_i = 
\left\{  
             \begin{array}{lr}  
             \min \big( W_i, \left( K_i + s_i \times (c-1) - p_i \right) \big) \ \ \  CONV, POOL  \\
             
             W_i \ \ \ \ \ \  \ \ \ \ \ \ \ \ \ \ \  \ \ \ \ \ \ \ \ \ \ \ \ \ \ \ \ \ \ \ \ \ \ \ \ \ \ \ \ \ \ \ \ \ \ \ \ \ \ \ \ \  FC \\
             
             (c_{d})_i \ \ \ \ \ \ \ \ \ \ \ \ \ \ \ \ \ \ \ \ \ \ \ \ \ \ \ \ \   CONCAT, ELTWISE \\
             \end{array}  
\right.  
\nonumber 
\end{equation} where $H_i$ and $W_i$ are the height and width of the output feature of node $i$, respectively. $K_i$, $s_i$, and $p_i$ are kernel size, stride, and padding size of node $i$, respectively, if node $i$ is convolutional or pooling layer.

According to the formula, the required input and expected output of each replicated block can be obtained. In order to improve computational parallelism and reduce the transmission between cores, other operations in DNN are divided into several cores according to the replication number of their predecessor convolutional layer.

\subsubsection{On-Chip Memory Reuse}

\begin{figure}[htbp]
  \centering
%   \vspace{-0.5cm} % 调整图片与上文的垂直距离
  \setlength{\abovecaptionskip}{-0.1cm} %调整标题上方的距离   
  \setlength{\belowcaptionskip}{-0.05cm} %调整标题下方的距离 	   
  \setlength{\abovedisplayskip}{0pt} %与上方展示内容的距离	
  \setlength{\belowdisplayskip}{0pt} %与下方展示内容的距离
  \includegraphics[width=0.7\linewidth]{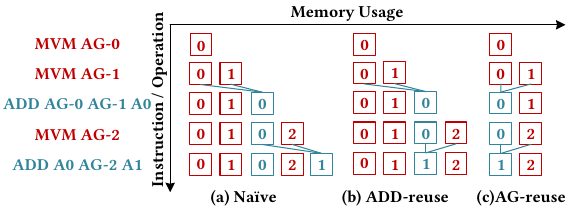}\\
  \caption{An example illustrating memory reuse optimization}
  \label{fig::memopt}
% \vspace{-0.8cm}
\end{figure}

% On-chip local memory resources are limited. 
\textcolor{black}{Since the capacity of on-chip local memory is limited and global memory access is expensive, efficient utilization of on-chip local memory is important to reduce the frequency of accessing global memory, improve executing efficiency and reduce power consumption.} Fig. \ref{fig::memopt} illustrates an example of memory optimization. For the naive method, we allocate new memory block for each operation. Most memory blocks are accessed once and will never be used again. So we use an ADD-reuse in Fig. \ref{fig::memopt}(b) to reuse the memory for data accumulation. But allocating memory blocks for each AG still causes a lot of waste. Therefore, we propose AG-reuse on the basis of ADD-reuse in Fig. \ref{fig::memopt}(c), which fully reuses the on-chip storage of AGs. 

% \vspace{-0.1cm}
\section{Evaluation}
\vspace{-0.25cm}

\subsection{Experiment Setup}
\vspace{-0.1cm}

\subsubsection{hardware characteristics}

\begin{figure*}
  \centering
%   \vspace{-1.2cm}
  \setlength{\abovecaptionskip}{-0.1cm} %调整标题上方的距离   
  \setlength{\belowcaptionskip}{-0.05cm} %调整标题下方的距离 	   
  \setlength{\abovedisplayskip}{0pt} %与上方展示内容的距离	
  \setlength{\belowdisplayskip}{0pt} %与下方展示内容的距离
  \includegraphics[width=0.85\linewidth]{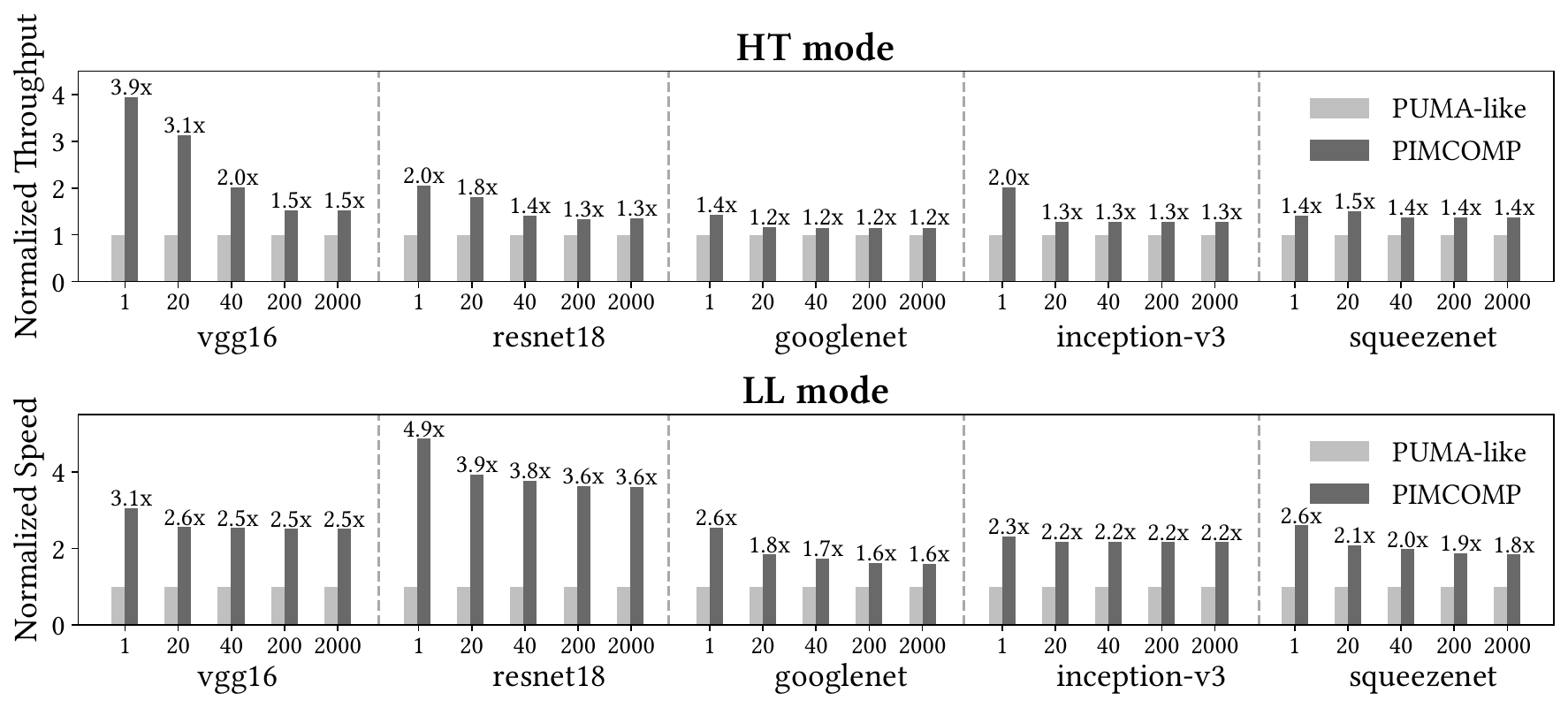}\\
%   \caption{Throughput and latency results with different parallelism (the maximum number of AGs that one core can process synchronously)}
 \caption{Throughput and latency results with different parallelism normalized to baseline}

  \label{fig::run}
\vspace{-0.1cm}
\end{figure*}

In order to evaluate the performance of PIMCOMP and facilitate fair comparison, we adopt PUMA architecture to instantiate our design. The PIM material is ReRAM, and NoC is selected as core connection implementation. Our evaluation adopts the same parameters as PUMA, including area and power information of ReRAM crossbars, VFUs, and control units. The ReRAM cell precision is 2-bit, and inputs, outputs and weights are 16-bit fixed-point numbers. Memory modules and routers are modeled by CACTI \cite{CACTI} and Orion 3.0 \cite{ORION} respectively, to get energy and area estimation. The detailed configurations of hardware are summarized in Table \ref{tab:hardware}.

\begin{table}[h]
\caption{Hardware Configurations}\vspace{-8pt}
\label{tab:hardware}
\small  % 调整字体
\resizebox{\columnwidth}{!} {
\begin{tabular}{ccccc}
\toprule
Component       & Parameters      & Specification & \begin{tabular}[c]{@{}l@{}}Power\\  ($mW$)\end{tabular} & \begin{tabular}[c]{@{}l@{}}Area \\ ($mm^2$)\end{tabular} \\ \midrule
PIMMU & \# crossbar    & 64            & 1221.76    & 0.77         \\
VFU             & \# per core    & 12            & 22.80      & 0.048        \\
Local Memory    & capacity       & 64 kB          & 18.00      & 0.085        \\
Control Unit    &   \textemdash       & \textemdash        & 8.00       & 0.11         \\
\textbf{Core}          & \textbf{\# per chip}    & \textbf{36}            & \textbf{1270.56}    & \textbf{1.01}         \\
Router          & ﬂit size       & 64            & 43.13      & 0.14         \\
Global Memory   & capacity       & 4 MB          & 257.72     & 2.42         \\
Hyper Transport  & link bandwidth & 6.40 GB/s      & 10.40 k    & 22.88        \\
\textbf{Chip}            &   \textemdash      &  \textemdash   & \textbf{56.79 \ k}     & \textbf{62.92}        \\ \bottomrule

\end{tabular}
% }
}
\vspace{-0.1cm}
\end{table}

\subsubsection{benchmarks and baselines}

% Our front end is able to parse the ONNX model to get a structured network description. 
We use both computationally intensive network (vgg16) and topologically complex networks (resnet18, squeezenet, googlenet and inception-v3) as benchmarks. We build a cycle-accurate simulator. It accepts the operation stream compiled by PIMCOMP, and can simulate the structure conflict and data dependency of MVM operation, the usage of on-chip local memory, the synchronization overhead of inter-core communication and can obtain energy consumption and area overhead.

As a comparison, we faithfully implement the PUMA dataflow under our framework and it is called a PUMA-like dataflow. According to  \cite{PUMA,PUMA-pre}, the purpose of node replicating is to balance the pipeline and a heuristic method is adopted to deal with core mapping. Since PUMA only support dataflow with pipeline granularity of inference (HT mode), we implement the LL mode for PUMA.

\vspace{-0.1cm}
\subsection{Experimental Result}
\vspace{-0.1cm}

\subsubsection{Throughput and Latency}

Fig. \ref{fig::run} shows the results of throughput and latency of PIMCOMP under different degrees of parallelism. The degree of parallelism here refers to how many AGs are allowed to calculate at the same time, limited by the user given on-chip bandwidth. According to the results, we can observe that with the increase of parallelism, the improvement of PIMCOMP gradually decreases. This is because the source of optimization is the gap between the actual performance and the ideal performance of the hardware. Still, PIMCOMP gains $1.6\times$ and $2.4\times$ improvements in throughput and latency, respectively, relative to PUMA.

In HT mode, the optimization effect of googlenet and squeezenet is limited. The main reason is that the MVM calculation pressure of these networks is light, so the time to access global memory and to process vector and memory operations become dominant. For computationally intensive tasks such as vgg16, PIMCOMP can achieve better optimization effect. In LL mode, the optimization effect of PIMCOMP is more significant because the node replicating method adopted by PUMA is not efficient enough.

\subsubsection{Energy}

\begin{figure}[htbp]
\vspace{-0.5cm}
  \centering
%   \vspace{-0.5cm} % 调整图片与上文的垂直距离
  \setlength{\abovecaptionskip}{-0.30cm} %调整标题上方的距离   
  \setlength{\belowcaptionskip}{-0.30cm} %调整标题下方的距离 	   
  \setlength{\abovedisplayskip}{0pt} %与上方展示内容的距离	
  \setlength{\belowdisplayskip}{0pt}
  %与下方展示内容的距离
  \includegraphics[width=0.98\linewidth]{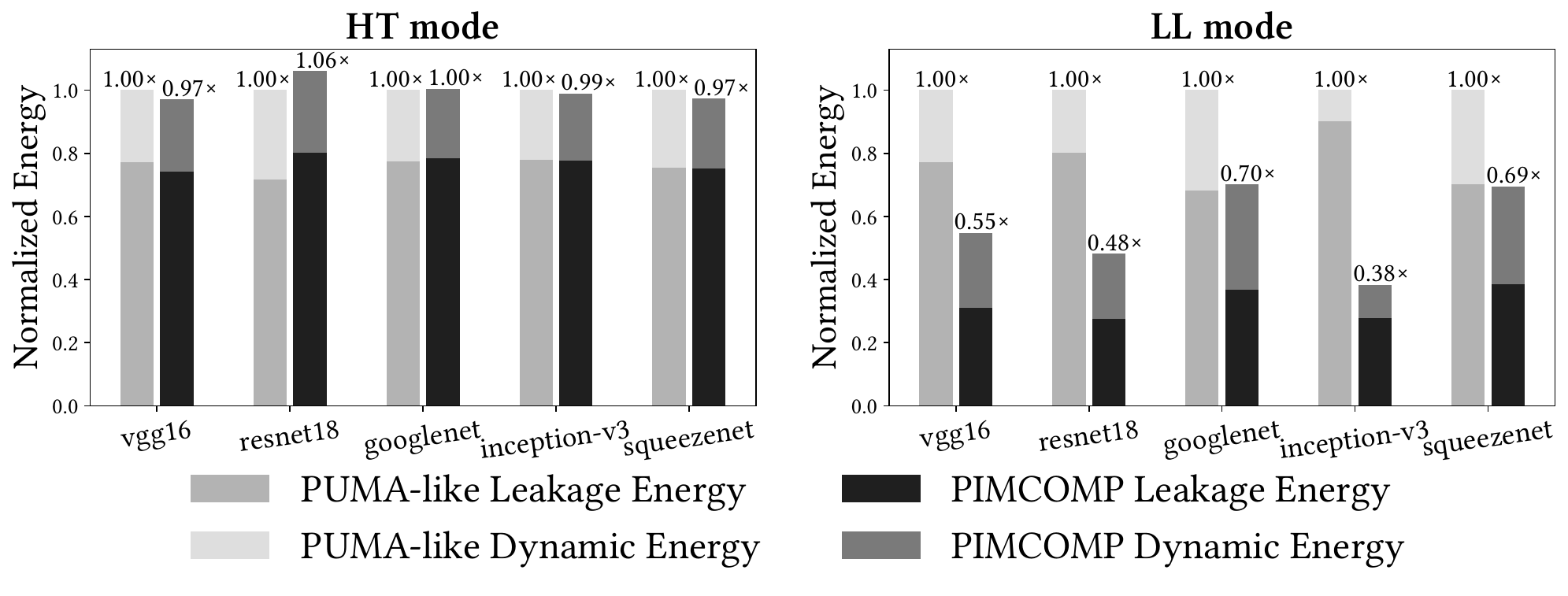}\\
  \caption{Energy breakdown for two compilation modes normalized to baseline}
  \label{fig::energy}
\vspace{-0.2cm}
\end{figure}

Fig. \ref{fig::energy} shows the energy evaluation results with parallelism degree $20$. The computational load of the same network is relatively fixed so dynamic energy of PUMA and PIMCOMP are close. The main difference between two compilation results is static energy consumption. In HT mode, PUMA allocates computation unevenly, causing some cores to run for a long time while others finish early. PIMCOMP ensures the computing tasks are evenly distributed. This results in a slight increase in static energy due to having more cores active, although PIMCOMP has a shorter overall runtime. In LL mode, there is data dependency between cores, and the active time of each core is related to the overall inference time. PIMCOMP is able to reduce static energy by $58.3\%$ by reducing overall runtime.

\subsubsection{Memory Usage}

\begin{figure}[htbp]
% \vspace{-0.4cm}
  \centering
%   \vspace{-0.3cm} % 调整图片与上文的垂直距离
  \setlength{\abovecaptionskip}{-0.15cm} %调整标题上方的距离   
  \setlength{\belowcaptionskip}{-0.1cm} %调整标题下方的距离 	   
  \setlength{\abovedisplayskip}{0pt} %与上方展示内容的距离	
  \setlength{\belowdisplayskip}{0pt}
  %与下方展示内容的距离
  \includegraphics[width=0.98\linewidth]{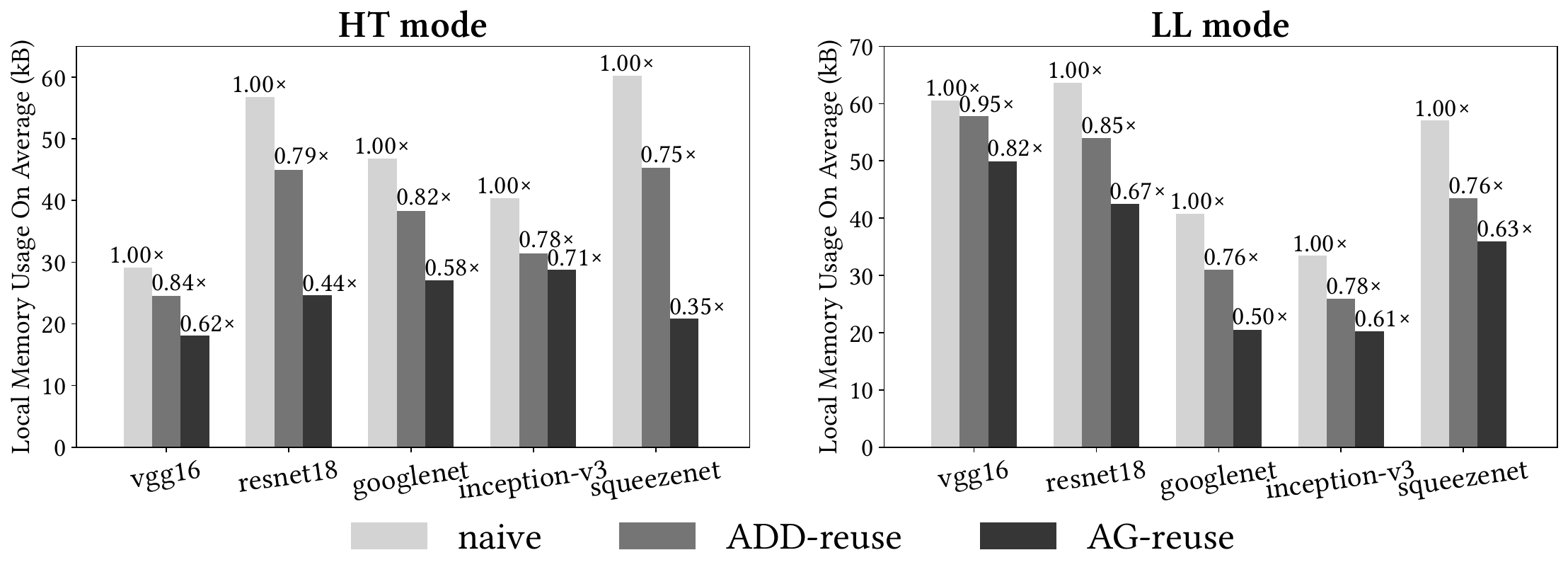}\\
  \caption{On-chip local memory usage with memory reuse optimization}
  \label{fig::mem}
% \vspace{-0.3cm}
\end{figure}

Fig. \ref{fig::mem} shows the effect of different memory reuse optimization. In the evaluation of HT mode, every core will transfer results back to global memory and load new input data into local memory after each AG performs 2 MVM  operations. In HT mode, if AG-reuse is adopted, the global memory access can be reduced by an average of $47.8\%$ compared with the naive method. This will result in faster inference speed and lower memory access energy consumption. In LL mode, the average local memory usage can be controlled within 64kB using AG-reuse optimization, which is in line with our architectural design.

\subsubsection{Compiling Time}

\begin{table}[!t]
\vspace{-0.3cm}
\caption{Compiling Time (second) For Benchmarks}\vspace{-8pt}
\label{tab:compile}
\small  % 调整字体
\resizebox{1.0\columnwidth}{!} {
\begin{tabular}{@{}ccccccccccc@{}}
\toprule
          & 
          \multicolumn{2}{c}{vgg16} & \multicolumn{2}{c}{resnet18} & \multicolumn{2}{c}{googlenet} & \multicolumn{2}{c}{squeezenet} & \multicolumn{2}{c}{inception\_v3} \\ 
          \cmidrule(lr){2-3}  \cmidrule(lr){4-5} \cmidrule(lr){6-7} \cmidrule(lr){8-9}
          \cmidrule(lr){10-11}
          & HT           & LL         & HT            & LL           & HT            & LL            & HT             & LL            & HT              & LL              \\ \midrule
Node Partitioning & 0.01         & 0.01       & 0.04          & 0.03         & 0.04          & 0.04          & 0.05           & 0.05          & 0.03            & 0.03            \\
Replicating+Mapping   & 8.93         & 1.80       & 12.39         & 6.35         & 12.90         & 8.10          & 12.04          & 7.43          & 12.88           & 8.76            \\
Dataflow Scheduling  & 1.62         & 6.67       & 0.54          & 4.39         & 0.64          & 5.44          & 1.08           & 32.72         & 0.80            & 20.78           \\
Total     & 10.56        & 8.48       & 12.96         & 10.78        & 13.57         & 13.58         & 13.17          & 40.21         & 13.71           & 29.57           \\ \bottomrule
\end{tabular}
}
\vspace{-0.6cm}
\end{table}

Table \ref{tab:compile} shows the running time of PIMCOMP. The population size is $100$ and the maximum iteration number is $200$ in the genetic algorithm. We can observe that weight replicating and core mapping take longer time in HT mode, while dataflow scheduling is time-consuming in LL mode. The overall compiling time is acceptable.

% \vspace{-0.5cm}
\vspace{-4pt}
\section{Conclusion}
\vspace{-0.1cm}

Previous research studies on NVM crossbar based PIM accelerator downplay several practical issues, such as parallelism consideration, weight replication selection and array mapping methods. In this work, we propose PIMCOMP, which has 4 general optimization stages to form a complete compilation toolchain for PIM accelerators. Evaluations show that our work outperforms a PUMA-like compiler on average in terms of performance and power consumption. PIMCOMP is orthogonal to other studies on PIM accelerators, and researchers can apply PIMCOMP to their  accelerators for further improvement.

\vspace{-4pt}
\bibliographystyle{ieeetr}
\bibliography{PIMCOM-Ref}

\begin{thebibliography}{10}

\bibitem{Diannao}
T.~Chen {\em et~al.}, ``Diannao: A small-footprint high-throughput accelerator
  for ubiquitous machine-learning,'' in {\em ASPLOS}, p.~269–284, 2014.

\bibitem{Eyeriss}
Y.-H. Chen {\em et~al.}, ``Eyeriss: A spatial architecture for energy-efficient
  dataflow for convolutional neural networks,'' in {\em ISCA}, pp.~367--379,
  2016.

\bibitem{MemoryWall}
S.-L. Lu {\em et~al.}, ``Scaling the “memory wall”: Designer track,'' in
  {\em ICCAD}, pp.~271--272, 2012.

\bibitem{ReRAMApp}
I.~Chakraborty {\em et~al.}, ``Resistive crossbars as approximate hardware
  building blocks for machine learning: Opportunities and challenges,'' {\em
  Proceedings of the IEEE}, vol.~108, no.~12, pp.~2276--2310, 2020.

\bibitem{ISAAC}
A.~Shafiee {\em et~al.}, ``Isaac: A convolutional neural network accelerator
  with in-situ analog arithmetic in crossbars,'' in {\em ISCA}, pp.~14--26,
  2016.

\bibitem{AtomLayer}
X.~Qiao {\em et~al.}, ``Atomlayer: A universal reram-based cnn accelerator with
  atomic layer computation,'' in {\em DAC}, pp.~1--6, 2018.

\bibitem{Brahms}
T.~Song {\em et~al.}, ``Brahms: Beyond conventional rram-based neural network
  accelerators using hybrid analog memory system,'' in {\em DAC},
  pp.~1033--1038, 2021.

\bibitem{MRAM}
S.~Jung {\em et~al.}, ``A crossbar array of magnetoresistive memory devices for
  in-memory computing,'' {\em Nature}, vol.~601, pp.~211--216, Jan 2022.

\bibitem{PCM2}
A.~Chen {\em et~al.}, ``Enabling high-performance dnn inference accelerators
  using non-volatile analog memory (invited),'' in {\em EDTM}, pp.~1--4, 2020.

\bibitem{PUMA}
A.~Ankit {\em et~al.}, ``Puma: A programmable ultra-efficient memristor-based
  accelerator for machine learning inference,'' in {\em ASPLOS}, p.~715–731,
  2019.

\bibitem{TVM}
T.~Chen {\em et~al.}, ``Tvm: An automated end-to-end optimizing compiler for
  deep learning,'' in {\em OSDI}, (USA), p.~579–594, 2018.

\bibitem{TC}
N.~Vasilache {\em et~al.}, ``Tensor comprehensions: Framework-agnostic
  high-performance machine learning abstractions,'' 2018.

\bibitem{MIXED}
Z.~Zhu {\em et~al.}, ``Mixed size crossbar based rram cnn accelerator with
  overlapped mapping method,'' in {\em ICCAD}, pp.~1--8, 2018.

\bibitem{InfoX}
Y.~He {\em et~al.}, ``Infox: An energy-efficient reram accelerator design with
  information-lossless low-bit adcs,'' in {\em DAC}, p.~97–102, 2022.

\bibitem{MNSIM2}
Z.~Zhu {\em et~al.}, ``Mnsim 2.0: A behavior-level modeling tool for
  memristor-based neuromorphic computing systems,'' in {\em GLSVLSI},
  p.~83–88, 2020.

\bibitem{CACTI}
R.~Balasubramonian {\em et~al.}, ``Cacti 7: New tools for interconnect
  exploration in innovative off-chip memories,'' {\em ACM Trans. Archit. Code
  Optim.}, vol.~14, jun 2017.

\bibitem{ORION}
A.~B. Kahng {\em et~al.}, ``Orion3.0: A comprehensive noc router estimation
  tool,'' {\em IEEE Embedded Systems Letters}, vol.~7, no.~2, pp.~41--45, 2015.

\bibitem{PUMA-pre}
J.~Ambrosi {\em et~al.}, ``Hardware-software co-design for an analog-digital
  accelerator for machine learning,'' in {\em ICRC}, pp.~1--13, 2018.

\end{thebibliography}

\end{document}